\documentclass[12pt]{article}
\usepackage{amsmath,amssymb,exscale}  
\input epsf

\def\gappeq{\mathrel{\rlap {\raise.5ex\hbox{$>$}}
{\lower.5ex\hbox{$\sim$}}}}
\def\permil{$\%\raise.20ex\hbox{$_0$}}
\def\lappeq{\mathrel{\rlap{\raise.5ex\hbox{$<$}}
{\lower.5ex\hbox{$\sim$}}}}

\begin{document}
\topmargin -1.0cm
\oddsidemargin -0.8cm
\evensidemargin -0.8cm
\pagestyle{empty}
\begin{flushright}
UNIL-IPT-00-26\\
hep-ph/0011105\\
November 2000
\end{flushright}
\vspace*{5mm}

\begin{center}

{\Large\bf Primordial magnetic fields from
inflation???\footnote{Contribution to the Proceedings of the Conference
on Cosmology and Astroparticle Physics CAPP2000, July
2000, Verbier, Switzerland.}}

%\vspace{1.0cm}

{\large Massimo Giovannini \footnote{Email:massimo.giovannini@ipt.unil.ch}
and Mikhail Shaposhnikov\footnote{Email: mikhail.shaposhnikov@ipt.unil.ch}}\\
\vspace{.6cm}
{\it {Institute of Theoretical Physics\\ University of Lausanne\\ 
CH-1015 Lausanne, Switzerland}}
\vspace{.4cm}
\end{center}

%\vspace{0.5cm}

\begin{abstract}
In this note we argue that the breaking of conformal invariance
because of the coupling of a charged scalar field to gravity is not
sufficient for the production of seed galactic magnetic fields during
inflation. 
\end{abstract}

Since the problem of long ranged magnetic fields is well known and is
thoroughly discussed in the literature (for a very recent review see
\cite{Grasso:2000wj}) we are not going to describe it here. 
We will instead concentrate on the recent controversy concerning the 
question of production of primordial seed magnetic fields from scalar
field fluctuations during inflation.

The original idea, due to Turner and Widrow \cite{Turner:1988bw}, can
be formulated as follows. While the coupling of electromagnetic field
to the metric and to the charged fields is conformally invariant, the
coupling of the charged  scalar field to gravity is not. Thus, vacuum
fluctuations of the charged scalar field can be amplified during
inflation over super-horizon scales, leading, potentially to non-trivial
correlations of the electric currents and charges over cosmologically
interesting distances. The  fluctuations of electric currents, in
their turn, may induce magnetic fields through Maxwell equations at
the corresponding scales. The role of the charged scalar field may be
played by the Higgs boson which couples to the hypercharge field
above the electroweak phase transition. The generated hypercharged
field is converted into ordinary magnetic field at the temperatures
of the order of electroweak scale. 

This suggestion was further investigated in \cite{Calzetta:1998ku}
for the standard electroweak theory with a conclusion that large
scale  magnetic fields can be indeed generated. These estimates were
challenged in \cite{Giovannini:2000dj}, where the computation similar
to \cite{Calzetta:1998ku} has been carried out, but with the adequate
treatment of kinetics of the hot plasma and realistic dependence of
electric conductivity on the temperature. Moreover, in
\cite{Giovannini:2000dj} the accurate expression for the amplified
currents has been derived without the use of  WKB approximation
(which has been assumed in \cite{Calzetta:1998ku}). The inclusion of
these effects changed the estimates  of \cite{Calzetta:1998ku} to a
level not sufficient for seeding of the galactic magnetic fields.

Very recently, a new proposal, also based on inflation, was put
forward \cite{Davis:2000zp} (see also \cite{Bassett:2000aw}). The
authors use the fact that during the exponential expansion of the
Universe the long ranged fluctuations of the charged scalar field are
amplified. This leads to the mass generation for the vector field
and, therefore, to the breaking of U(1) symmetry. The inflationary
phase is then  replaced by a  radiation dominated stage  of
expansion, and the scalar field fluctuations relax to zero, leading
to the restoration of the U(1) symmetry. In Ref.\cite{Davis:2000zp}
it was claimed that the abrupt change of the mass of the photon at
the end of the inflationary stage results in production of magnetic
fields which may be sufficiently large to seed the galactic magnetic
fields, and in ref. \cite{Bassett:2000aw} it was claimed that the 
coherent oscillations of the charged scalar field during preheating
lead to the similar effect.

In the conformal time coordinate  $\tau$, the authors of
\cite{Davis:2000zp,Bassett:2000aw} write  the evolution equation of
the  fluctuations of the gauge fields as:
\begin{equation}
\left( \partial_\tau^2 + \vec{k}^2 + e^2 a^2 \langle \rho^2\rangle
\right) A_i(\vec{k},\tau)=0~,
\label{main}
\end{equation}
where $e$ is the charge of scalars, $a$ is the scale factor, $\langle
\rho^2\rangle$ is the magnitude of the scalar field fluctuations 
(computed in the unitary gauge), and $A_i(\vec{k},\tau)$ is the Fourier
harmonic of the vector field. Taking a step-function approximation
for the change of the scalar field fluctuations the authors of
\cite{Davis:2000zp} found that the resulting spectrum of the
generated field strengths is approximately $B_l \propto l^{-1}$,
where $l$ is the relevant coherence scale. In ref.\cite{Bassett:2000aw}
an oscillating behaviour of the scalar field was assumed.

Our main objection to the analysis of
refs. \cite{Davis:2000zp,Bassett:2000aw} is that Eq. (\ref{main})
disregards dissipative effects that are crucially important for the
generation of magnetic fields and charge density fluctuations. Though
the importance of dissipation was mentioned in refs.
\cite{Davis:2000zp,Bassett:2000aw}, it was not taken into account in
computation of amplification of gauge field fluctuations.  While in
de Sitter stage of expansion one can assume that the equations of
motion in the vacuum are adequate, they are certainly not true in
radiation dominated epoch containing a plasma of charged particles.
In particular, these equations are not correct during the preheating
stage, where charged scalars are copiously produced simply because of
their self-interaction. They are also not true right after inflation,
since the change of the gravitational background leads to creation of
charged scalar particles. In the media, containing charged particles,
the change of vector potential gives rise to electric fields that
accelerate particles, leading, ultimately, to  damping effects. A
phenomenological way\footnote{A more rigorous treatment, based on the
Landau-Vlasov kinetic equation, is discussed in
\cite{Giovannini:2000dj}.} to incorporate the damping effects is to
add to the left-hand side of Eq. (\ref{main}) a damping term
\cite{Calzetta:1998ku} 
\begin{equation}  
\sigma a\partial_{\tau}A_i(\vec{k},\tau) ~,
\end{equation}  
where  $\sigma$ is a plasma conductivity. The lower limit on the
conductivity can be established simply by the counting of the number
of scalar particles produced during inflation and is given by
equations (5.8) and (B.23) of Ref. \cite{Giovannini:2000dj}, which
lead to $\sigma \sim \frac{H}{\alpha}$, where $H$ is the Hubble
constant during inflation and $\alpha$ is the fine structure
constant. One can check that that the damping term is greater than
the one with the second derivative, so that the damping (rather than
amplification!) of the gauge field fluctuations is given by
\begin{equation}
A_i(\vec{k},\tau)=\exp \left(-\int_{\tau_0}^\tau d\tau \frac{e^2 a^2 \langle
\rho^2\rangle}{\sigma a}\right) A_i(\vec{k},\tau_0)~,
\end{equation}
rather than what was found in Refs.\cite{Davis:2000zp,Bassett:2000aw}. Here
$A_i(\vec{k},\tau_0)$ is the typical gauge field fluctuation at the
end of inflation.  
  
The effect of the conductivity is crucial exactly because the
fluctuations of the massive vector field {\em do not grow} during
inflation and in fact they are smaller than the ones of a free
massless vector field that couples to gravity in a
conformally-invariant way. This follows from the analysis of the
correlation  functions of the magnetic fields during the 
de Sitter phase, based on eq. (\ref{main}),
\begin{equation}
\langle B_{i}(\vec{x},\tau) B_{j} (\vec{y},\tau) \rangle =
\int d^3 k\,\,\tau\,\,P_{ij}(k) H^{(1)}_{\mu}(k\tau) 
\,\,H^{(2)}_{\mu}(k\tau) e^{i \vec{k} \cdot (\vec{x} - \vec{y})},
\end{equation}
where 
\begin{equation}
P_{i j}(k) = \frac{k^2}{32 ~\pi^2} \biggl( \delta_{ij} - 
\frac{k_i k_j}{k^2}\biggr), 
\end{equation}
and $H_{\mu}^{(1)}(k\tau) = {H_{\mu}^{(2)}}^{\ast}(k\tau)$ are the 
Hankel functions of first and second order.
As a consequence the spectrum of magnetic fields 
(in terms of the length scale $l$) is of the form 
 $B_l \propto l^{-\nu}$ where $\nu = 2 +
{\cal O}(\frac{e^2}{\lambda}) >2 $. 
Thus, the amplitude of the magnetic field, created by this
mechanism,  is too small on cosmological distances.

We thank Tomislav Prokopec and Ola T\"ornkvist for helpful
discussions.

\end{document}